\PassOptionsToPackage{unicode}{hyperref}
\PassOptionsToPackage{hyphens}{url}
\documentclass[12pt]{article}

\def\spacingset#1{\renewcommand{\baselinestretch}%
{#1}\small\normalsize} \spacingset{1}

\usepackage{amsmath,amssymb}
\usepackage{lmodern}
\usepackage{ifxetex,ifluatex}
\usepackage{amsmath}
\usepackage{graphicx,psfrag,epsf}
\usepackage{enumerate}
\usepackage{url} 
\usepackage[T1]{fontenc}
\usepackage[utf8]{inputenc}
\usepackage[page]{appendix}
\usepackage[margin=1in]{geometry}
\usepackage{color}
\usepackage{fancyvrb}
\usepackage{upquote}
\usepackage[]{microtype}
\UseMicrotypeSet[protrusion]{basicmath} 
\usepackage{xcolor}
\usepackage{booktabs}
\usepackage{longtable}
\usepackage{array}
\usepackage{multirow}
\usepackage{wrapfig}
\usepackage{float}
\usepackage{colortbl}
\usepackage{pdflscape}
\usepackage{tabu}
\usepackage{threeparttable}
\usepackage{threeparttablex}
\usepackage[normalem]{ulem}
\usepackage{makecell}
\usepackage{framed}
\usepackage{hyperref}
\usepackage{authblk}

\DefineVerbatimEnvironment{Highlighting}{Verbatim}{commandchars=\\\{\}}
\usepackage{framed}
\definecolor{shadecolor}{RGB}{248,248,248}
\newenvironment{Shaded}{\begin{snugshade}}{\end{snugshade}}

\newcommand{\AttributeTok}[1]{\textcolor[rgb]{0.77,0.63,0.00}{#1}}

\newcommand{\CommentTok}[1]{\textcolor[rgb]{0.56,0.35,0.01}{\textit{#1}}}

\newcommand{\ConstantTok}[1]{\textcolor[rgb]{0.00,0.00,0.00}{#1}}
\newcommand{\ControlFlowTok}[1]{\textcolor[rgb]{0.13,0.29,0.53}{\textbf{#1}}}

\newcommand{\DecValTok}[1]{\textcolor[rgb]{0.00,0.00,0.81}{#1}}

\newcommand{\FunctionTok}[1]{\textcolor[rgb]{0.00,0.00,0.00}{#1}}

\newcommand{\NormalTok}[1]{#1}

\newcommand{\OtherTok}[1]{\textcolor[rgb]{0.56,0.35,0.01}{#1}}

\newcommand{\SpecialCharTok}[1]{\textcolor[rgb]{0.00,0.00,0.00}{#1}}

\newcommand{\StringTok}[1]{\textcolor[rgb]{0.31,0.60,0.02}{#1}}

\usepackage{graphicx}
\makeatletter
\def\maxwidth{\ifdim\Gin@nat@width>\linewidth\linewidth\else\Gin@nat@width\fi}
\def\maxheight{\ifdim\Gin@nat@height>\textheight\textheight\else\Gin@nat@height\fi}
\makeatother
\setkeys{Gin}{width=\maxwidth,height=\maxheight,keepaspectratio}
\makeatletter
\def\fps@figure{htbp}
\makeatother
\providecommand{\tightlist}{%
  \setlength{\itemsep}{0pt}\setlength{\parskip}{0pt}}
\usepackage{soul}
\usepackage{todonotes}
\usepackage{booktabs}
\usepackage{longtable}
\usepackage{array}
\usepackage{multirow}
\usepackage{wrapfig}
\usepackage{float}
\usepackage{colortbl}
\usepackage{pdflscape}
\usepackage{tabu}
\usepackage{threeparttable}
\usepackage{threeparttablex}
\usepackage[normalem]{ulem}
\usepackage{makecell}
\usepackage{xcolor}
\ifluatex
  \usepackage{selnolig}  
\fi
\usepackage[]{natbib}
\newlength{\cslhangindent}
\newlength{\csllabelwidth}
\newenvironment{CSLReferences}[2] 
 {
  \setlength{\parindent}{0pt}
  \ifodd #1 \everypar{\setlength{\hangindent}{\cslhangindent}}\ignorespaces\fi
  \ifnum #2 > 0
  \setlength{\parskip}{#2\baselineskip}
  \fi
 }%
 {}
\usepackage{calc}

\title{Rethinking the Funding Line at the Swiss National Science
Foundation: Bayesian Ranking and Lottery}
\author[1,2,*]{Rachel Heyard}
\author[1]{Manuela Ott}
\author[3]{Georgia Salanti}
\author[3,4]{Matthias Egger}
\affil[1]{\small Data Team, Swiss National Science Foundation, Bern, Switzerland}
\affil[2]{\small Center for Reproducible Science, University of Zurich, Zurich, Switzerland}
\affil[3]{\small Institute of Social and Preventive Medicine, University of Bern, Bern, Switzerland}
\affil[4]{\small Population Health Sciences, Bristol Medical School, University of Bristol, Bristol, UK}
\affil[*]{Corresponding author: Rachel Heyard, rachel.heyard@uzh.ch}
\date{}

\begin{document}
\maketitle
\begin{abstract}
\noindent Funding agencies rely on peer review and expert panels to
select the research deserving funding. Peer review has limitations,
including bias against risky proposals or interdisciplinary research.
The inter-rater reliability between reviewers and panels is low,
particularly for proposals near the funding line. Funding agencies are
also increasingly acknowledging the role of chance. The Swiss National
Science Foundation (SNSF) introduced a lottery for proposals in the
middle group of good but not excellent proposals. In this article, we
introduce a Bayesian hierarchical model for the evaluation process. To
rank the proposals, we estimate their expected ranks (ER), which
incorporates both the magnitude and uncertainty of the estimated
differences between proposals. A provisional funding line is defined
based on ER and budget. The ER and its credible interval are used to
identify proposals with similar quality and credible intervals that
overlap with the provisional funding line. These proposals are entered
into a lottery. We illustrate the approach for two SNSF grant schemes in
career and project funding. We argue that the method could reduce bias
in the evaluation process. R code, data and other materials for this
article are available online.
\end{abstract}

\noindent \emph{Keywords:} grant peer review, expected rank, posterior
mean, Bayesian ranking, funding line, modified lottery \vfill 

\hypertarget{introduction}{%
\section{Introduction}\label{introduction}}

Public research funding is limited and highly competitive. Not every
grant proposal can be funded, even if the idea is worthwhile and the
research group highly qualified. Funding agencies face the challenge of
selecting the proposals or researchers that merit support among all
proposals submitted to a call for applications. Funders generally rely
on expert peer review for determining which projects deserve to be
funded \citep{harman_management_1998}. For example, in the UK, over 95\%
of medical research funding was allocated based on peer review
\citep{guthrie_what_2018}. At the Swiss National Science Foundation
(SNSF), external experts first assess the proposals which are then
reviewed and discussed by the responsible panel, taking into account the
external peer review \citep{severin_gender_2020}.

The evidence base on the effectiveness of peer review is limited
\citep{guthrie_measuring_2019}, but it is clear that peer review of
grant proposals has several limitations. Bias against highly innovative
and risky proposals is well documented \citep{guthrie_what_2018} and
probably exacerbated by low success rates, potentially leading to
``conservative, short-term thinking in applicants, reviewers, and
funders'' \citep{alberts_rescuing_2014}. There is also evidence of bias
against highly interdisciplinary projects. An analysis of the Australian
Research Council's Discovery Programme showed that the greater the
degree of interdisciplinarity, the lower the success rate
\citep{bromham_interdisciplinary_2016}. The data on gender bias are
mixed, but women tend to have lower publication rates and lower success
rates for high-status research awards than men
\citep{kaatz_threats_2014, van_der_lee_gender_2015}.

Several studies have shown considerable disagreement between individual
peer reviewers assessing the same proposal, with kappa statistics
typically well below 0.50
\citep{cicchetti_reliability_1993, cole_chance_1981, fogelholm_panel_2012, guthrie_measuring_2019}.
A similar situation is observed at the level of evaluation panels. For
example, a study of the Finnish Academy compared the assessments by two
expert panels reviewing the same grant proposals
\citep{fogelholm_panel_2012}. The kappa for the consolidated panel score
of the two panels after the discussion was 0.23. Interestingly, the same
kappa was obtained when using the mean of the scores from the external
reviewers, indicating that panel discussions did not improve the
evaluation's consistency. The low inter-rater reliability means that a
research proposal's funding decision will partly depend on the peer
reviewers and the responsible panel, and therefore on chance.

The wisdom of a system for research funding decisions that depends
partly on chance has been questioned over many years
\citep{cole_chance_1981, mayo_peering_2006}. More recently,
\citet{fang_research_2016} argued that the current grant allocation
system is ``in essence a lottery without the benefits of being random''
and that the role of chance should be explicitly acknowledged. They
proposed a modified lottery where excellent research proposals are
identified based on peer review and those funded within a given budget
are selected at random. The Health Research Council of New Zealand
\citep{liu_acceptability_2020}, the
\citet{volkswagen_foundation_experiment_2017} and recently the
\citet{austrian_research_fund_1000_2020} are funders that have applied
lotteries, with a focus on transformative research or unconventional
research ideas. The SNSF introduced a modified lottery for its junior
fellowship scheme, focusing on proposals near the funding line
\citep{adam_science_2019, bieri_how_2020}.

The definition of the funding line is central in this context. Many
funders rank proposals based on the review scores' simple averages to
draw a funding line \citep{liaw_lucy_peer_2017}. Averages are
understandable to all stakeholders but not optimal for ranking.
Proposals around the funding line will often have similar or even
identical scores. Because the number of reviewers or panel members is
limited, the statistical evidence that two adjacent scores above and
below the funding line are different is often weak.
\citet{kaplan_sample_2008} have shown that unrealistic numbers of
reviewers (100 reviewers or more) per proposal would be required to
detect smaller differences between average scores reliably. The same
point has been made in the context of the ranking of baseball players
\citep{berger_bayesian-approach_1988}.

In this paper, we show how Bayesian ranking (BR) methods
\citep{laird_empirical_1989} can be used to define funding lines and
identify applications to be entered into a modified lottery. We will
describe the BR methodology with its expected rank and other statistics,
formulate recommendations on its use to support funding decisions, and
finally, apply the method to two grant schemes of the SNSF. Mandated by
the government, the SNSF is Switzerland's foremost funding agency,
supporting scientific research in all disciplines.

\hypertarget{methodology}{%
\section{Methodology}\label{methodology}}

\hypertarget{a-bayesian-hierarchical-model-for-the-evaluation-process}{%
\subsection{A Bayesian hierarchical model for the evaluation
process}\label{a-bayesian-hierarchical-model-for-the-evaluation-process}}

Let us assume a setting where \(n\) research proposals are being
submitted for funding. The proposals are graded for their quality using
a score on a 6-point interval scale (from 1, poor to 6, outstanding). A
research proposal \(i\) is evaluated by up to \(m\) distinct evaluators
/ assessors; so that \(y_{ij}\), with \(i\) in \(\{1, \dots, n\}\),
\(j\) in \(\{1, \dots, m\}\), is the score given to proposal \(i\) by
assessor \(j\).

Conditional on proposal and assessor effects, the observations \(y_{ij}\) are assumed to be normally distributed with a mean that depends on the proposal and assessor. The parameters of
interest for ranking are the true underlying proposal effects
\(\theta_i\).

\begin{eqnarray}
y_{ij} \ | \ \theta_i, \lambda_{ij}  & \sim & N(\bar{y} + \theta_i + \lambda_{ij}, \sigma^2) \label{eq:final-model} \\
\theta_i & \sim & N(0, \tau_{\theta}^2) \nonumber \\
\lambda_{ij} & \sim & N(\nu_j, \tau_{\lambda}^2), \nonumber
\end{eqnarray} where \(\bar{y}\) is the overall mean score of all
proposals. The model residuals are the measurement error for assessor
\(j\) on proposal \(i\). In the model above, \(\tau_{\theta}^2\) can be
interpreted as the total variability of the proposals around the mean.
To account for the tendency of assessors to be more or less strict in
their scoring, we add a parameter \(\lambda_{ij}\). We assume that this
parameter follows a normal distribution with mean \(\nu_j\) and variance
\(\tau^2_{\lambda}\). The assessor specific mean \(\nu_j\) accounts for
different scoring habits of the assessors. A stricter assessor, compared
to all other assessors, will have a negative \(\nu_j\); a more generous
assessor will have a positive \(\nu_j\).
Note that $y_{ij}$ is centered around the overall mean score only if the expected value of $\nu_j$ is $0$ (which holds under our prior specification for $\nu_j$ in Section \ref{sec:estimation}).
The estimation of these parameters will be discussed later in Section
\ref{sec:estimation}.

Alternatively, a Bayesian hierarchical model for ordinal data can be
used. Following the work of \cite{Johnson2008} and \cite{Cao2010} we
assume that the proposals have an underlying continuous quality trait on
which the true ranking is based. The assessors essentially estimate this
trait and compare it to fixed cutoffs to obtain the ordinal score. The
cutoffs are assumed to be the same for each assessor. \(\theta_i\) will
be the underlying quality trait for proposal \(i\), which is estimated
by assessor \(j\) to be \(x_{ij}\). \(x_{ij}\) is a latent, non-observed
variable, while \(y_{ij}\) is the observed score given by assessor \(j\)
using certain cutoffs \(c_k\). Let us assume that \(K\) is the number of
levels of the ordinal score, \(k \in \{1, \dots, K \}\), \(c_o = -\inf\)
and \(c_K = \inf\), then the model is defined as follows:

\begin{eqnarray}
x_{ij} & \sim & N(\bar{y} + \theta_i + \lambda_{ij}, \sigma^2)  \label{eq:ordinal-model} \\
y_{ij} & = & k \iff c_{k-1} < x_{ij} \leq c_k \nonumber \\
\theta_i & \sim & N(0, \tau_{\theta}^2) \nonumber \\
\lambda_{ij} & \sim & N(\nu_j, \tau_{\lambda}^2). \nonumber
\end{eqnarray}

The interpretation of the remaining parameters in model
\eqref{eq:ordinal-model} stays the same as discussed for model
\eqref{eq:final-model}.

Until now we assumed that the residual variance of the normally
distributed scores is the same for all proposals \(i\), namely
\(\sigma^2\) in equations \eqref{eq:final-model} and
\eqref{eq:ordinal-model}. It is possible however that if the average
score of a proposal is very low or high, then the individual scores tend
to be less variable than for a proposal with an average in the middle
range. The Bayesian hierarchical models presented above can be extended
to heterogeneous residual variances by following the mixed-effects
location scale models introduced in \cite{Hedeker2008}. We simply
introduce proposal-specific residual variances \(\sigma_i^2\) and model
these variances as a function of the average scores per proposal. We
also include one more effect to account for additional (\textit{i.e.}
beyond the variation explained by the average scores) between-proposal
variation in the residual variances of the scores. More precisely, we
extend the hierarchical model \eqref{eq:final-model} as follows:
\begin{eqnarray}
y_{ij} \ | \ \theta_i, \lambda_{ij}  & \sim & N(\bar{y} + \theta_i + \lambda_{ij}, \sigma_i^2), \label{eq:hetero-model} \\
\theta_i & \sim & N(0, \tau_{\theta}^2) \nonumber \\
\lambda_{ij} & \sim & N(\nu_j, \tau_{\lambda}^2), \nonumber \\
\sigma_i^2 & = & \exp(\alpha + \beta \log(\bar{y}_i) + \omega_i) \nonumber \\
\omega_i & \sim & N(0, \tau_{\omega}^2), \nonumber
\end{eqnarray} where \(\bar{y}_i\) is the mean of the scores for
proposal \(i\). Note that if we assign a normal prior to \(\omega_i\),
then the variances \(\sigma_i\) have a log-normal distribution
\citep{Hedeker2008}. The same extension can directly be applied on the
ordinal outcome model in \eqref{eq:ordinal-model}.

If the proposals are evaluated in different sub-panels and the results
of all sub-panels have to be merged into a final ranking, the likelihood
can be adapted to include a panel effect \(\delta_k\): \begin{eqnarray}
y_{ijk} \ | \ \theta_i, \lambda_{ij}, \delta_k & \sim & N(\bar{y} + \theta_i + \lambda_{ij} + \delta_k, \sigma^2) \label{eq:merged-model} \\
\theta_i & \sim & N(0, \tau_{\theta}^2) \nonumber \\
\lambda_{ij} & \sim & N(\nu_j, \tau_{\lambda}^2) \nonumber\\ 
\delta_k & \sim & N ( 0, \tau_{\delta}^2), \nonumber
\end{eqnarray} where \(k\) refers to the panel. The mean of the prior
distribution on \(\delta_k\) is set to \(0\), because we assume the
panels being comparable and therefore unbiased (this assumption can of
course be relaxed, for example by allowing for a panel-specific mean
different from \(0\)). The ordinal and location scale model definitions
in \eqref{eq:ordinal-model} and \eqref{eq:hetero-model} could be
extended in a similar fashion to allow for different panels.
Note that in the use case where each assessor evaluates and grades only one proposal, the model would need to be simplified by deleting the assessor-specific effect and variance terms. In general, the ranking methodology described in the next sections can be applied to other Bayesian hierarchical models as long as the individual proposal quality and its uncertainty are estimated.

\hypertarget{ranking-grant-proposals}{%
\subsection{Ranking grant proposals}\label{ranking-grant-proposals}}

The key parameter on which we will base the ranking are the true
proposal effects, respectively the true quality traits, \(\theta_i\) in
the models defined above. Simply ranking proposals based on the means of
the posterior distribution of the \(\theta_i\)'s may be misleading
because it does not account for the uncertainty reflected by the
posterior variances of the \(\theta_i\)'s. \citet{laird_empirical_1989}
introduce the expected rank (\(\mbox{ER}_i\)) as the expectation of the
rank of \(\theta_i\):

\begin{eqnarray}
\mbox{ER}_i & = & \mbox{E}(\mbox{rank}(\theta_i)) \label{eq:er} \\
& = & \sum_{k = 1}^n \mbox{Pr}(\theta_i \leq \theta_k) \nonumber \\
& = & 1 + \sum_{i \neq k}\mbox{Pr}(\theta_i \leq \theta_k), \mbox{ because Pr}(\theta_i \leq \theta_i) = 1. \nonumber
\end{eqnarray}

Note that ER incorporates the magnitude \emph{and} the uncertainty of
the estimated difference between proposals. Scaling ER to a percentage
(from \(0\%\) to \(100\%\)) facilitates interpretation
\citep{van_houwelingen_empirical_2020}:
\(\mbox{PCER}_i = 100 \times (\mbox{ER}_i - 0.5)/n\). The
\emph{percentile based on ER} (PCER) is independent of the number of
competing proposals and is interpreted as the probability that proposal
\(i\) is of worse quality compared to a randomly selected proposal. A
related quantity, the \emph{surface under the cumulative ranking}
(SUCRA) line, summarizes all rank probabilities
\citep{salanti_graphical_2011}. It is based on the probability that
proposal \(i\) is ranked on the \(m\)-th place, denoted by
\(\mbox{Pr}(i = m)\). Then, these probabilities can be plotted against
the possible ranks, \(m = 1, \dots, n\), in a so-called rankogram. The
cumulative probabilities \(\mbox{cum}_{im}\) of proposal \(i\) being
among the \(m\) best proposals are summed up to form the SUCRA of
proposal \(i\):
\(\mbox{SUCRA}_i = \frac{\sum_{m = 1}^{n -1} \mbox{cum}_{im}}{n-1}\).
The higher the SUCRA value, the higher the likelihood of a proposal
being in the top ranks. As SUCRA is between 0 and 1, it can be
interpreted as the fraction of competing proposals that the proposal
\(i\) `beats' in the ranking. As shown in Appendix \ref{app:ersucra},
the SUCRA can directly be transformed to ER:
\(\mbox{ER}_i = n - (n - 1) \cdot \mbox{SUCRA}_i\). In order to assess
how reliable the votes of all the assessors were the intra-class
correlation coefficient (ICC) \citep{Shrout1979} can be computed.

In the following, we refer to Bayesian ranking (BR) as the ranking based
on the ER.

\hypertarget{estimation-and-implementation}{%
\subsection{\texorpdfstring{Estimation and implementation
\label{sec:estimation}}{Estimation and implementation }}\label{estimation-and-implementation}}

We pursue a fully Bayesian approach in JAGS (Just Another Gibbs Sampler)
using the \texttt{rjags} interface in \texttt{R}
\citep{plummer_package_2019}. Therefore, we need to define priors on the
parameters described in the models \eqref{eq:final-model},
\eqref{eq:ordinal-model}, \eqref{eq:hetero-model} and
\eqref{eq:merged-model}, respectively.

Since, in our case studies, the scores \(y_{ij}\) all lie in the
interval \([1,6]\), we can derive a worst-case upper bound for the
variance of the \(\theta_i\)'s and \(\lambda_{ij}\)'s by assuming a
uniform distribution on this interval. Apart from this upper bound, we
do not have much prior information on the variability of the
\(\theta_i\)'s and \(\lambda_{ij}\)'s. Therefore, as suggested by
\citet{gelman_prior_2006}, we use a uniform prior on \((0,2]\) for
\(\tau_{\theta}\), \(\tau_{\lambda}\), and \(\tau_{\delta}\) if the
evaluations are done in sub-panels. Similarly, we fix a uniform prior on
\((0,2]\) on \(\sigma\): \begin{eqnarray}
\tau_{\theta}, \tau_{\lambda}, \tau_{\delta}, \sigma & \sim &  U_{(0,2]} \nonumber \\
\nu_j & \sim & N(0, 0.5^2). \nonumber
\end{eqnarray} The parameter summarizing the assessor behavior,
\(\nu_j\), will follow a normal distribution around 0, with a small
variance (\(\sim 0.5^2\)), i.e.~about 95\% of the assessors are assumed
to have a assessor-specific mean in {[}-1, 1{]}.

For model \eqref{eq:ordinal-model} an additional prior has to be set on
the cutoffs for the ordinal scale. The quantities \(c_k\) denote the
common cutoffs for all assessors. A uniform prior on the real line is
defined on those cutoffs, with the constraint \(c_k < c_{k+1}\)
\citep{Cao2010}. In our case studies, we will be using \(K = 6\) since
the scale used in the evaluation is a six-point scale.

For model \eqref{eq:hetero-model}, we need to additionally specify
hyper-priors for the parameters \(\alpha\), \(\beta\) and
\(\tau_{\omega}\). We assume \begin{eqnarray}
\alpha & \sim & N(0, 10^2), \nonumber \\
\beta & \sim & N(0, 10^2), \nonumber \\
\tau_{\omega} & \sim & U(0, 10]. \nonumber
\end{eqnarray}

We implemented the presented methodology in \texttt{R} (see package
\href{https://snsf-data.github.io/ERforResearch/}{\texttt{ERforResearch}}
available on github, the package documentation and vignette as well as
the
\href{https://snsf-data.github.io/ERpaper-online-supplement/index.html}{online supplement}
of this paper).

\hypertarget{computational-aspects}{%
\subsubsection{\texorpdfstring{Computational aspects
\label{sec:comp}}{Computational aspects }}\label{computational-aspects}}

Whenever MCMC methods are used, convergence diagnostics need to be
considered. We decided to do visual inspections of the trace plots and
computed the \(\widehat{R}\) values of the Gelman-Rubin convergence
diagnostic \citep{Gelman1992} of the most important parameters.
Additionally we integrated a convergence test in the
implementation, that first uses the user specified number of iterations,
burn-in and adaptations of the JAGS algorithm. Then, the \(\widehat{R}\)
values of all the relevant parameters are computed. If the maximum value
of those \(\widehat{R}\) values is larger than the threshold of \(1.1\)
recommended by \citet{Gelman2014} the latter numbers (\texttt{n.burnin},
\texttt{n.apdapt} and \texttt{n.burnin}) are increased until the model
converges (as defined by having all \(\widehat{R}\) values \(\leq1.1\)).
However, the algorithm will stop after a maximum number of iterations
(default one million) and return a warning message. More information on
convergence diagnostics as well as the definition of the JAGS model
and hyper-prior sensitivity can be found in Appendix
\ref{app:conv} and in Sections 3 \and 4 of the
\href{https://snsf-data.github.io/ERpaper-online-supplement/index.html}{online supplement}.

\hypertarget{strategy-for-ranking-proposals-and-drawing-the-funding-line}{%
\subsection{Strategy for ranking proposals and drawing the funding
line}\label{strategy-for-ranking-proposals-and-drawing-the-funding-line}}

To rank the competing proposals we can use the following steps:

\begin{enumerate}
\def\labelenumi{(\arabic{enumi})}
\tightlist
\item
  Ranking proposals based on the means of the posterior distribution of
  the \(\theta_i\)'s. This strategy incorporates the uncertainty and
  variation from the evaluation process, but ignores the uncertainty
  induced by the posterior mean's estimation.
  Additionally, this is not a truly comparative ranking where all proposals are compared with every other proposal.
\item
  Using the expected ranks ER\(_i\) (or the equivalent PCER\(_i\) and
  SUCRA\(_i\)) to compare proposals instead. This approach incorporates
  all sources of variation that can be quantified.
\end{enumerate}

We start by plotting the ERs together with their 50\% credible interval
(CrI). Note that the 50\% CrI was a political choice, to ensure small
random selection groups (see Discussion Section \ref{sec:disc}). A
provisional funding line (FL) can be defined by simply funding the
best-ranked \emph{x} proposals. \emph{x} is generally determined by the
available budget. This FL is then exactly defined at the ER of the last
fundable proposal. The provisional funding line helps to see the bigger
picture. Are there clusters of proposals? Is there a group of proposals
with credible intervals of the rank that overlap with the provisional
FL? The latter group should be considered for inclusion into a lottery
where the proposals that can be funded are selected at random. If there
is enough funding to fund all proposals in the lottery, no random
selection element is needed. On the other hand, a clear distance between
the ERs of non-funded proposals and the provisional FL, meaning no
overlap in their 50\% CrI with the provisional FL, suggests that the
proposals can be easily separated with respect to their estimated
quality and no random selection is needed. This strategy is what we
refer to as Bayesian ranking.

\hypertarget{case-studies}{%
\section{Case studies}\label{case-studies}}

In this section, we will use the approach to simulate recommendations
for two funding instruments: the \emph{Postdoc.Mobility} fellowship for
early career researchers and the SNSF \emph{Project Funding} scheme for
established investigators. For all examples in the case studies we first
use the continuous outcome model as described in \eqref{eq:final-model}
and \eqref{eq:merged-model}, and then compare the results with the
results obtained from the ordinal model described in
\eqref{eq:ordinal-model}. Model \eqref{eq:hetero-model} will be used in
the simulation study, see Section \ref{sec:sim}.

\hypertarget{postdoc.mobility-pm-funding-scheme}{%
\subsection{Postdoc.Mobility (PM) funding
scheme}\label{postdoc.mobility-pm-funding-scheme}}

Junior researchers who recently defended their PhD and wish to pursue an
academic career can apply for a fellowship, which will allow them to
spend two years in a research group abroad. In a first step, each
proposal is allocated in one of five panels (Humanities; Social
Sciences; Science, Technology, Engineering and Mathematics (STEM);
Medicine; and Biology) and attributed to a referee and a co-referee who
will evaluate the proposal in detail. Both of the referees then score
the proposal on a six-point scale (from 1, poor, to 6, outstanding).
Based on these scores, all the proposals of a panel are triaged into
three groups: ``fund'', ``discuss'' (in panel) and ``reject''. The
middle group's proposals are further discussed in the respective panels
to take the final funding decisions. The panel members score each of the
proposals and agree on a funding line, with or without the usage of a
lottery for some proposals. We used the data from the February 2020
call. Due to the pandemic, the meeting was remote, and the scoring
independent: panel members did not know how fellow members scored the
proposals. Table \ref{tab:number_applications_pm} summarizes the number
of proposals in the different ``panel discussion'' groups together with
the number of proposals that can still be funded and the size of the
panel. Note that each panel had its own assessors, \textit{i.e.} panel
members.

\begin{table}[!h]

\caption{\label{tab:number-applications-pm}Characteristics of panels and submissions for the 
               February 2020 call for junior fellowships at the Swiss National 
               Science Foundation. \label{tab:number_applications_pm}}
\centering
\begin{tabular}[t]{l>{\raggedleft\arraybackslash}p{3cm}>{\raggedleft\arraybackslash}p{3cm}>{\raggedleft\arraybackslash}p{3cm}>{\raggedleft\arraybackslash}p{3cm}}
\toprule
  & N. of panel members & Total N. of submissions & N. of proposals discussed & N. of fundable proposals\\
\midrule
Humanities & 14 & 23 & 11 & 4\\
Social Sciences & 14 & 38 & 18 & 7\\
Biology & 26 & 35 & 18 & 8\\
Medicine & 17 & 35 & 14 & 7\\
STEM & 29 & 50 & 18 & 6\\
\bottomrule
\end{tabular}
\end{table}

Table \ref{tab:icc_tab} shows the intra-class correlation coefficients
representing the assessor reliability calculated for each of the five
panels. Additionally, the 95\% confidence interval is provided (computed
with the function \texttt{ICC} in the R package \texttt{psych}). A more
detailed figure presenting all the votes of the panel members for each
of the proposals discussed can be found in
\href{https://snsf-data.github.io/ERpaper-online-supplement/index.html}{the online supplement (Figure 1.1)}.
In all five panels, the assessor reliability can be classified from poor
to fair.

\begin{table}[!h]

\caption{\label{tab:ICC table}The intra-class correlation coefficients together with their
        95\% confidence intervals of the different panels.
        \label{tab:icc_tab}}
\centering
\begin{tabular}[t]{ll}
\toprule
  & ICC (95\% CI)\\
\midrule
Humanities & 0.33 (0.18; 0.58)\\
Social Sciences & 0.4 (0.27; 0.58)\\
Biology & 0.5 (0.37; 0.67)\\
Medicine & 0.43 (0.28; 0.63)\\
STEM & 0.31 (0.2; 0.47)\\
\bottomrule
\end{tabular}
\end{table}

\begin{figure}
\centering
\includegraphics{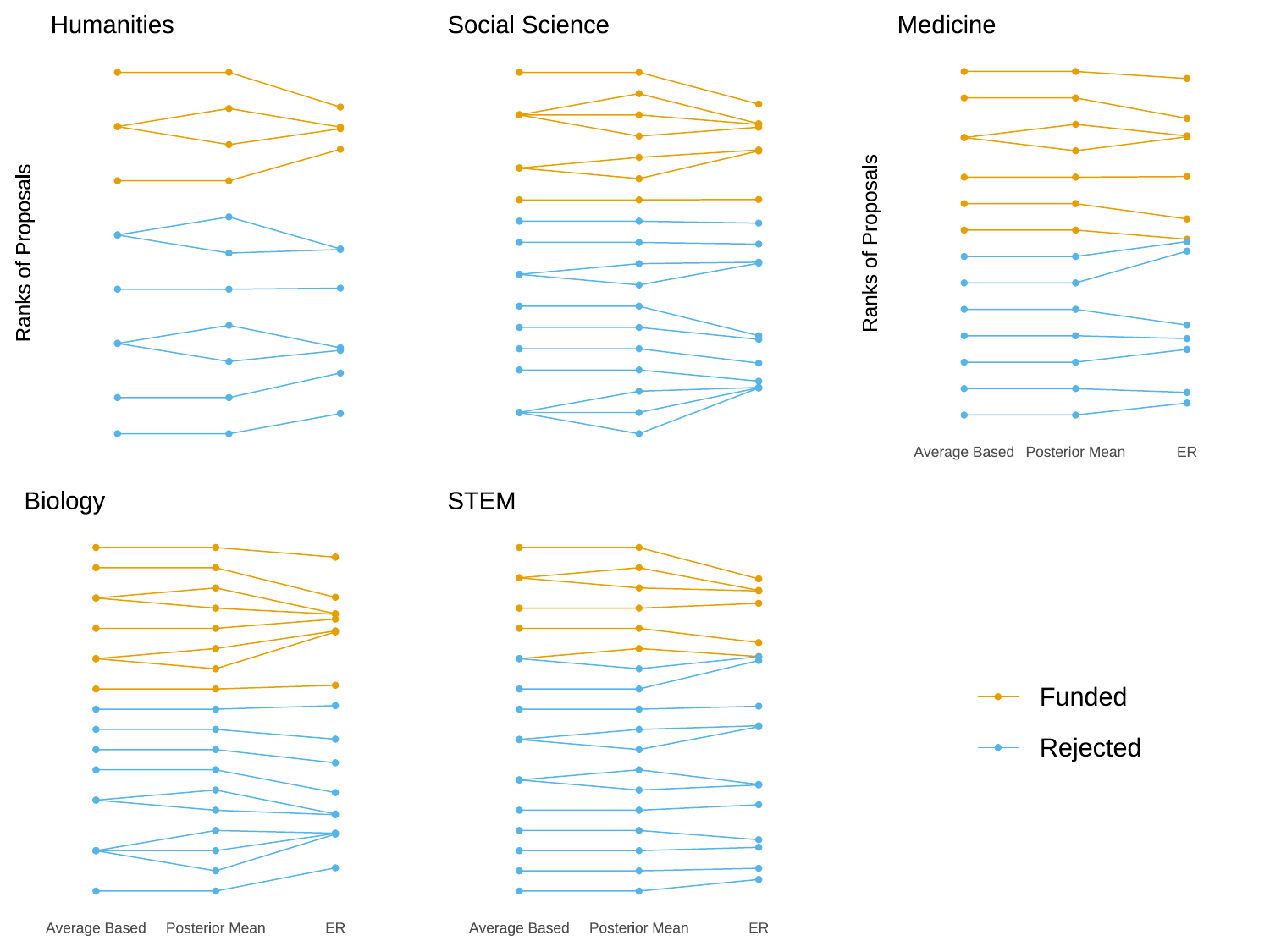}
\caption{Junior fellowship proposals ranked based on simple averages
(points on the left), the posterior means (middle) and the ER (points on
the right), by the evaluation panel. Note that rank 1, displayed on the
top, is the best rank. The color indicates the proposals funded if the
best proposals based on the ER and constrained by the number of fundable
proposals are funded.\label{fig:ER-plot-pm2020}}
\end{figure}

For the funding decision, the \(\theta_i\)'s defined in
\eqref{eq:final-model} are of primary interest. We calculated the
distribution of the rank of the \(\theta_i\)'s, and obtain the ER as the
posterior mean of this distribution. Figure \ref{fig:ER-plot-pm2020}
shows the different ways of ranking the proposals, for all panels
seperately. The points in the left column show the ranking based on the
simple averages (if two proposals have the same average grade, they are
on the same rank). Next, indicated by the middle column, the proposals
are ranked based on the posterior means of the \(\theta_i\)'s (posterior
mean ranking). Finally, the points in the right column show the expected
rank. A provisional funding line represented by the change of color is
defined by simply funding the \(x\) first ranked proposals, where \(x\)
is the number of fundable proposals in Table
\ref{tab:number_applications_pm}. This kind of presentation is solely
used for illustrative purposes, and not for funding recommendations. The
representation shows how the expected ranks relate to the ranking of the
proposals based on their average score.

\begin{figure}
\centering
\includegraphics{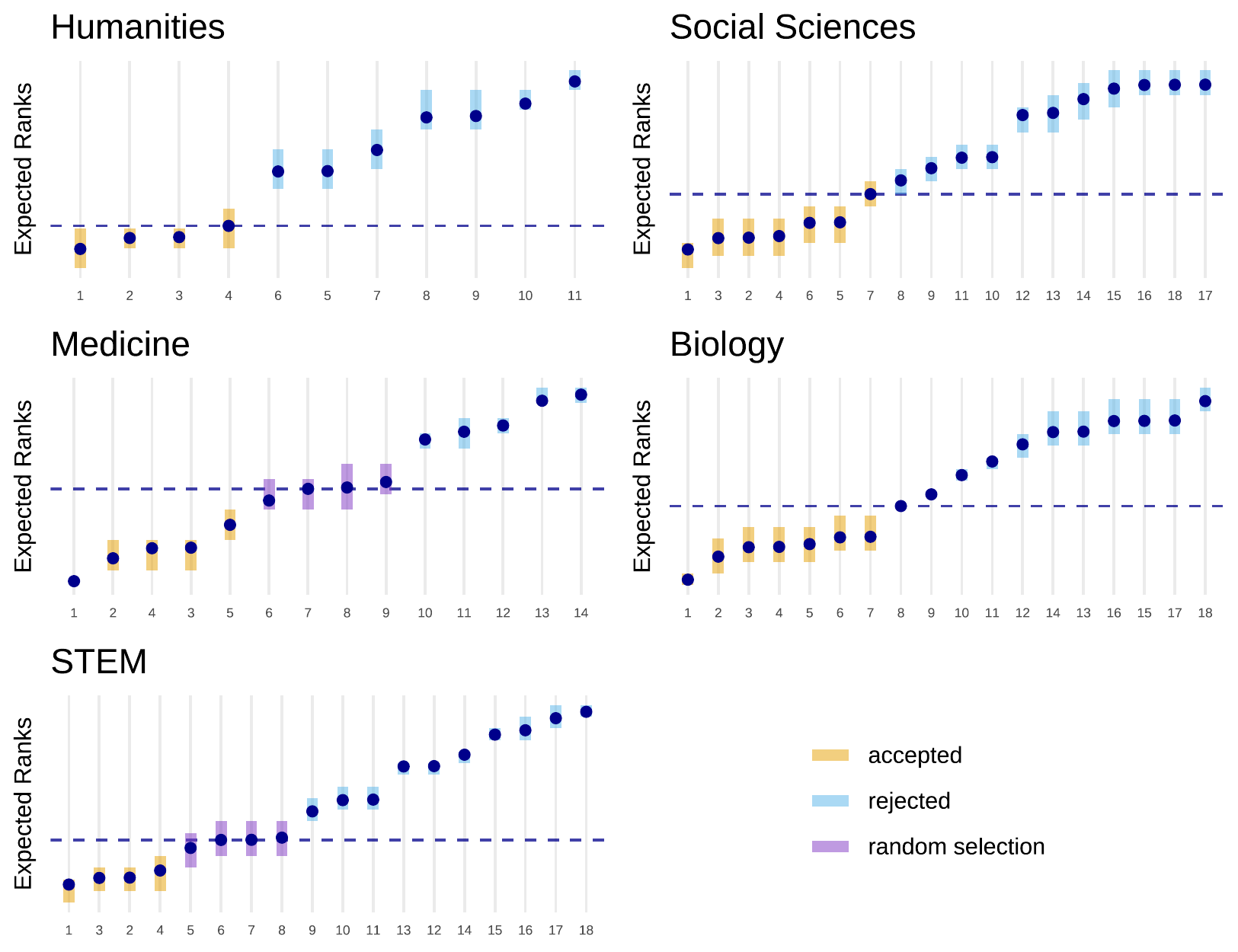}
\caption{Bayesian Ranking: The expected rank as point estimates
(posterior mean), together with 50\% credible intervals (colored boxes).
The dashed blue line is the provisional funding line, \(i.e.\) the ER of
the last fundable proposal. The color code indicates the final group the
proposal is in: accepted or rejected proposals, or random
selection/lottery group. \label{fig:rank_credible_intervals_pm2020}}
\end{figure}

Figure \ref{fig:rank_credible_intervals_pm2020} plots the ERs of the
same proposals together with their 50\% credible intervals and the
provisional funding line (\(=\) the ER of the last fundable proposal).
This presentation facilitates identifying proposals that cluster around
the funding line; i.e.~the proposals that might be included in a lottery
(or random selection group). The methodology supports the following
decisions:

\begin{itemize}
\tightlist
\item
  Humanities panel: The four best ranked proposals are funded, the
  remaining seven are rejected, no random selection.
\item
  Social Sciences: The seven best ranked proposals are funded, the
  eleven worst ranked proposals are rejected, no random selection.
\item
  Medicine: The five best ranked proposals are funded, the five worst
  ranked proposals are rejected. Two proposals are randomly selected for
  funding among the four proposals ranked as sixth to ninth.
\item
  Biology: The eight best ranked proposals are funded, the ten worst
  ranked proposals are rejected, no random selection.
\item
  STEM: The four best ranked proposals are funded, the ten worst ranked
  proposals are rejected. Two proposals are randomly selected for
  funding among the four proposals ranked as fifth to eighth.
\end{itemize}

\begin{figure}
\centering
\includegraphics{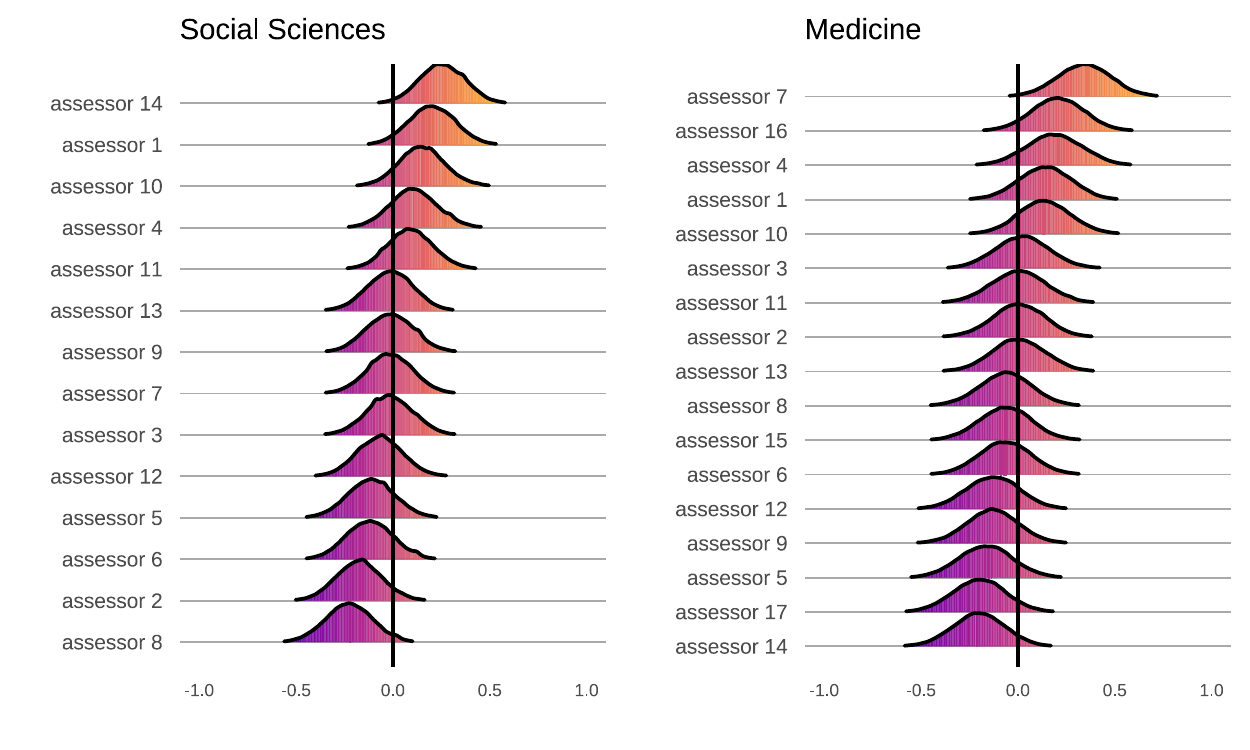}
\caption{Posterior distributions of the assessor-specific means
\(\nu_j\) in the Social Sciences and Medicine panels. Each of the 14
assessors voted on up to 18 proposals, unless they had a conflict of
interest or other reasons to be absent during the vote. Note that the 14
assessors in the Social Sciences panel were not the same as the ones in
the Medicine panel. \label{fig:voter_behav_ER_pm_2020}}
\end{figure}

Samples from the posterior distributions of all the parameters in the
Bayesian hierarchical model can be extracted from the JAGS model. This
allows the funders to better understand the evaluation process. As a
reminder, parameter \(\nu_j\) summarises the behavior of panel member
\(j\). The more \(\nu_j\) is negative, the stricter the scoring behavior
of assessor \(j\) compared to the remaining panel members. This also
means that the stricter grades from assessor \(j\) are corrected more,
because they comply with their usual behavior. Figure
\ref{fig:voter_behav_ER_pm_2020} shows the posterior distributions of
the \(\nu_j\)'s for the Social Sciences and Medicine panels. Note that
we only present these two panels, because they are small enough to allow
interpretation. The illustration of the remaining panels can be found in
\href{https://snsf-data.github.io/ERpaper-online-supplement/index.html}{Figure 1.4 in the online supplement}.
In the Social Sciences, assessor 8 is a more critical panel member,
whereas assessor 14 gives, on average, the highest scores. Also in the
Medicine panel, the distributions for the different assessors are quite
different. This illustrates how important it is to account for assessor
effects.

Additionally, the posterior means of the variation of the proposal
effects, \(\tau_{\theta}^2\), for each panel with 90\% CrI can be
extracted: 0.14 with {[}0.06, 0.34{]} for the Humanities, 0.13 with
{[}0.07, 0.25{]} for the Social Sciences, 0.18 with {[}0.1, 0.32{]} for
the Biology Panel, 0.17 with {[}0.08, 0.34{]} for the Medicine panel and
0.17 with {[}0.09, 0.31{]} for the STEM panel. The same information can
be retrieved for the variation of the assessor effect
\(\tau_{\lambda}^2\): 0.09 with {[}0, 0.25{]} for the Humanities, 0.07
with {[}0, 0.18{]} for the Social Sciences, 0.06 with {[}0, 0.16{]} for
the Biology Panel, 0.07 with {[}0, 0.2{]} for the Medicine panel and
0.13 with {[}0, 0.34{]} for the STEM panel.

Further Figures, representing the rankograms and SUCRAs, the PCER and
the ranking using the posterior mean of \(\theta_i\) and their 50\% CrI
can be found
\href{snsf-data.github.io/ERpaper-online-supplement}{in Figures 1.5 to 1.7 in the online supplementary material}.
For all the results of the panels presented above we used the convergence test described in Section \ref{sec:comp} to ensure convergence of all chains.

Similar results are found when using the ordinal model defined in
\eqref{eq:ordinal-model}, see
\href{snsf-data.github.io/ERpaper-online-supplement}{Figure 1.8 in the online supplementary material}.
\href{snsf-data.github.io/ERpaper-online-supplement}{Figure 1.9 in the online supplement}
shows the Bland-Altman plot \citep{Bland1999} representing the agreement
between the rankings based on the ordinal and continuous models, defined
in equations \eqref{eq:final-model} and \eqref{eq:ordinal-model}
respectively.

\hypertarget{project-funding}{%
\subsection{\texorpdfstring{Project Funding
\label{subsec:pf}}{Project Funding }}\label{project-funding}}

Project Funding is the SNSF's most important funding instrument. Project
grants support blue-sky research of the applicant's choice. We analyzed
the proposals submitted to the April 2020 Call to the Mathematics,
Natural and Engineering Sciences (MINT) division. Overall, the division
evaluated 353 grant proposals. The evaluation was done in four
sub-panels of the same size (nine members) and a similar number of
international and female members. Each panel member evaluated all
proposals (unless they had a conflict of interest), and each panel
defined its own funding line aiming at a similar (\(\sim 30\%\)) success
rate. Table \ref{tab:summarising} shows the total number of proposals,
fundable proposals and average scores (from 1, poor, to 6, outstanding).
Average scores were highest in panel 3 and lowest in panel 4.

\begin{table}[!h]

\caption{\label{tab:summarising-table}Summarising statistics in the four panels of the 
  Mathematics, Natural and Engineering Sciences division in Project Funding.
  \label{tab:summarising}}
\centering
\begin{tabular}[t]{l>{\raggedleft\arraybackslash}p{3cm}>{\raggedleft\arraybackslash}p{3cm}>{\raggedleft\arraybackslash}p{3cm}>{\raggedleft\arraybackslash}p{3cm}}
\toprule
Panel & N. of discussed proposals & N. of fundable proposals & Average score & Average score in top 30\%\\
\midrule
Panel one & 87 & 26 & 3.84 & 5.01\\
Panel two & 92 & 28 & 3.81 & 5.02\\
Panel three & 86 & 26 & 3.97 & 5.24\\
Panel four & 88 & 26 & 3.73 & 5.04\\
\bottomrule
\end{tabular}
\end{table}

Table \ref{tab:icc_tab_mint} shows the ICC (with 95\% confidence
interval).
\href{snsf-data.github.io/ERpaper-online-supplemen}{Figure 2.1 in the online supplement}
presents all the votes of the panel members in each sub-panel for each
of the proposals. Compared to the previous Postdoc.Mobility case study,
the intra-class correlation coefficients are higher and can be
classified as good. One reason for the higher reliability here is that
all Project Funding proposals are included - also the ones with low/high
average scores, for which the votes are usually less variable. In
contrast, for Postdoc.Mobility we only looked at the proposals with
average scores in the middle range.

\begin{table}[!h]

\caption{\label{tab:ICC-table-Mint}The intra-class correlation coefficients 
        together with their 95\% confidence intervals of the different
        sub-panels. The overall reliability in the those panels can be
        classified as good.\label{tab:icc_tab_mint}}
\centering
\begin{tabular}[t]{ll}
\toprule
  & ICC (95\% CI)\\
\midrule
Panel one & 0.82 (0.78; 0.86)\\
Panel two & 0.85 (0.82; 0.89)\\
Panel three & 0.85 (0.81; 0.88)\\
Panel four & 0.82 (0.78; 0.86)\\
\bottomrule
\end{tabular}
\end{table}

\begin{figure}
\includegraphics[width=62.22in]{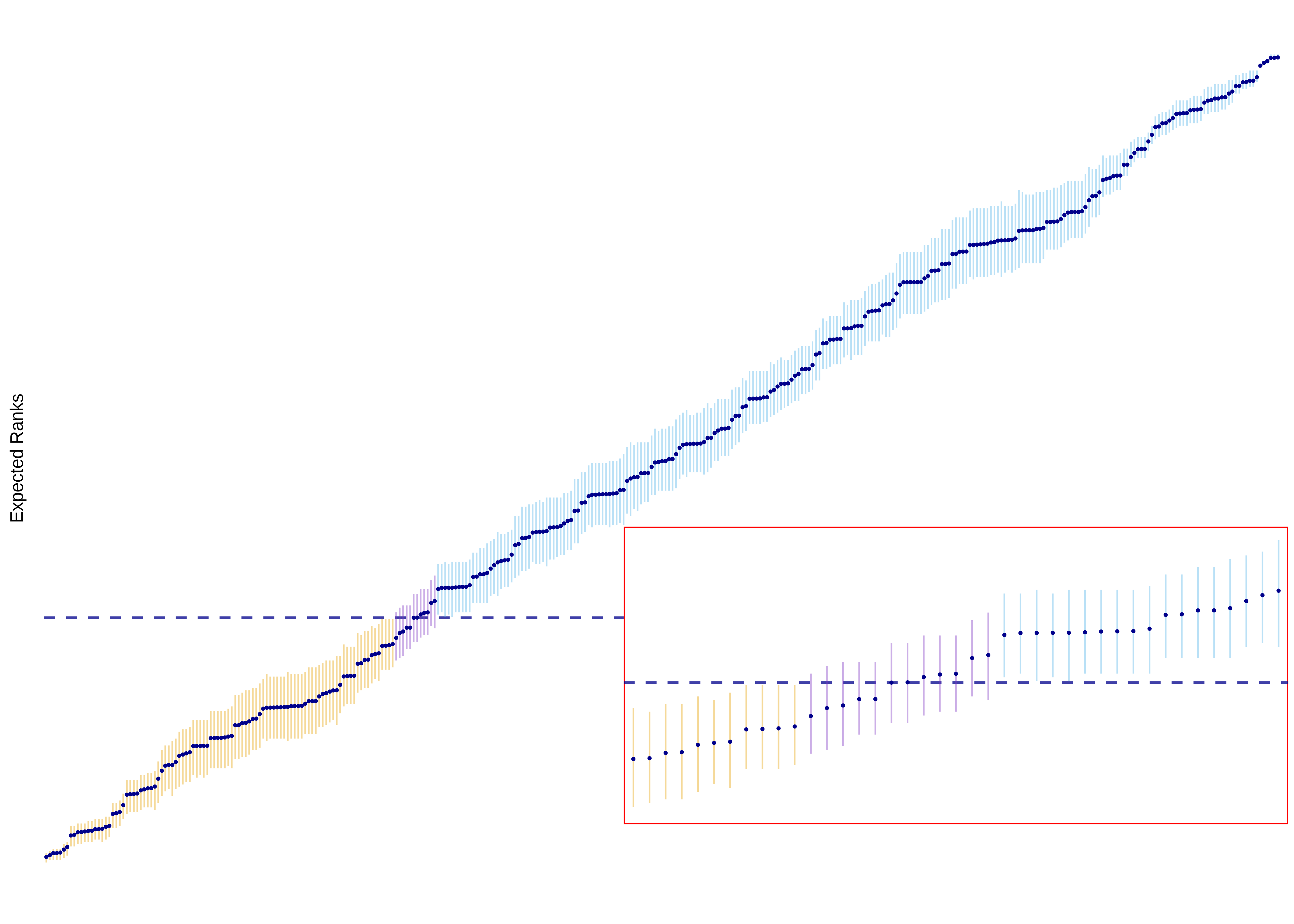} \caption{Bayesian Ranking: Proposals ordered from the best ER (bottom left, rank 1) to the worst (top right, rank 352) together with their 50\% credible intervals. A provisional funding line to ensure a 30\% success rate is drawn (blue dashed line). The proposals are arranged in three groups: funded (orange), random selection (violet) and rejected (blue).\label{fig:rank_mint_all}}\label{fig:mint-er-plot-all}
\end{figure}

In this case study, we illustrate the BR with an underlying model
defined in \eqref{eq:merged-model}, where an overall ranking is aimed
for, while the evaluation was done in sub-panels. Figure
\ref{fig:rank_mint_all} shows the ER ordered from the best-ranked
proposal (bottom left) to the worst (top right) together with their 50\%
credible intervals. The provisional funding line is defined as the ER of
the last fundable proposal: the 106th (30\% of 353) best ranked,
according to its ER. Zooming in on the provisional funding line shows
the cluster of proposals with similar quality and 50\% credible
intervals overlapping with the funding line. These proposals may be
included in the lottery/random selection group.

As for the Postdoc.Mobility evaluations, we can also estimate the
variation of the proposal, assessor and panel effects (posterior mean
with 90\% CrI): 1.13 with {[}0.99, 1.28{]} for \(\tau_{\theta}^2\), 0.09
with {[}0, 0.2{]} for \(\tau_{\lambda}^2\) and 0.01 with {[}0, 0.12{]}
for \(\tau_{\delta}^2\). A more detailed analysis of this case study can
be found in the
\href{snsf-data.github.io/ERpaper-online-supplement}{online supplement}
(especially, also the results if using model \eqref{eq:final-model} on
the sub-panels seperately).
For all the results presented above we again used the convergence test described in Section \ref{sec:comp} to ensure convergence of all chains.

Similar results are found when using a Bayesian hierarchical model that
explicitly accounts for the ordinal nature of the scores; see
\href{snsf-data.github.io/ERpaper-online-supplement}{Figure 2.5 in the online supplement}.
\href{snsf-data.github.io/ERpaper-online-supplement}{Figure 2.6 in the online supplement}
shows the Bland-Altman plot; the agreement between the BR with an
ordinal outcome model and the BR with a continuous outcome model.
However, even though both models yield similar results, it is hard to
say which modelling approach performs better without knowing exactly
which proposals should have been funded or rejected. We therefore
compare the performance of different models in a simulation study, where
the true ranks are known.

\hypertarget{simulation-study}{%
\section{\texorpdfstring{Simulation Study
\label{sec:sim}}{Simulation Study }}\label{simulation-study}}

We conducted a simulation study to compare our proposed Bayesian ranking
method (with continuous and ordinal outcomes, homogeneous and
heterogeneous residuals) to the currently used procedure, \textit{i.e.}
the ranking based on averages. We closely follow \cite{Cao2010} to
simulate the true ranks. We suppose that our panel is composed of
\(J = 10\) members, who assess \(I = 50\) proposals on an ordinal scale
from 1 to 6 (where 6 is the best score and 1 is the lowest). Note that
in Cao et al.'s simulation study a five-point scale was used; to be as
close as possible to our current scenario we use a six-point scale
instead. The data is simulated from a model that is different from the
one assumed for the computation of the Bayesian ranking. More
specifically, we start with the ordinal outcome model in
\eqref{eq:ordinal-model} and apply the following adaptations. For
\(j = 1, \dots, 10\), \(\alpha_j \sim \mbox{Unif}(-0.5, 0.5)\) and for
\(j = 1, \dots, 9\), the variation will be defined as
\(\sigma_j \sim \mbox{Unif}(0, 0.5)\) and \(\sigma_{10} = 0.9\). We
assume \(\tau_{\theta} = 1\), so that \(\theta_i \sim N(0,1)\) for
\(i = 1, \dots, I\). Finally, the latent variable
\(x_{ij} \sim N(\alpha_j + \theta_i, \sigma_j^2)\) and the final grade
\(y_{ij}\) is simulated with the cutoffs
\(c_1 = -1, c_2 = -0.5, c_3 = 0, c_4 = 0.5, c_5 = 1\). To mimic the
possibility of conflicts of interest or abstentions in the data, we
first draw a random number \(M \in [1, 100]\) for the number of missing
scores and then randomly add \(M\) missing values into the data matrix.
We replicate these simulation steps
$N = 500$ times to achieve a small enough Monte-Carlo Error (see below).
The code to simulate the data can be found in Appendix \ref{app:code}.
To estimate the Bayesian hierarchical models, we use 10'000 iterations,
5'000 burn-in and adaptation iterations and four chains.

The true ranks are computed using the simulated true proposal effects
(\(\theta^{\mbox{\tiny true}}_i, i = 1, \dots, J\)):
\(R_i^{\mbox{\tiny true}}(\mathbf{\theta}) = \mbox{rank}(\theta^{\mbox{\tiny true}}_i) = \sum_{l = 1}^I \mbox{I}(\theta^{\mbox{\tiny true}}_i > \theta^{\mbox{\tiny true}}_l).\)
Then, in order to compare the different approaches to each other, and
understand which method best estimated the true rank we will use the
mean squared error (MSE) loss:
\[\mbox{MSE} = \frac{1}{I} \sum_i(R^{\mbox{\tiny true}}_i(\mathbf{\theta}) - R_i^{est})^2,\]
with \(R_i^{est}\) being the estimated rank; the rank of proposal \(i\)
given by the procedure that is tested.

The approaches compared here are the ranking based on the average, the
Bayesian Ranking (BR) with an underlying normal outcome model and
homogeneous residuals (as described in \eqref{eq:final-model}), the BR
with an ordinal outcome model and homogeneous residuals (as described in
\eqref{eq:ordinal-model}), the BR with a normal outcome model and
heterogeneous residuals, and finally the BR with an underlying ordinal
outcome model and heterogeneous residuals (both adapted from
\eqref{eq:hetero-model}). Table \ref{tab:tab_sim_res} compares the mean
squared error of the different ranking approaches.
Following the suggestion from \cite{Morris2019} we additionally calculated the Monte Carlo standard error (SE) for the MSE as follows

\[\mbox{Monte Carlo SE} = \frac{\sum_i\left ((R^{\mbox{\tiny true}}_i(\mathbf{\theta}) - R_i^{est})^2 - \mbox{MSE} \right)^2}{N (N -1)}.\]
As mentioned in \cite{Cao2010} the MSE informs us on the ability of the
procedure to accurately rank all the proposals. The best performing
method is the BR with underlying ordinal model and heterogeneous
residuals; the model closest to the data generating process. However,
the simple BR with underlying normal outcome model and homogeneous
residuals outperforms the remaining approaches. Hence, applying the
Occam razor principle, favoring simpler models to complex models, we
would suggest to use the BR approach as defined in model
\eqref{eq:final-model}. The reason why this simpler model performs so
well is most likely due to the fact that for the ordinal outcome model
(and the heterogeneous residuals) an additional latent variable, the
cutoffs and other parameters have to be estimated.

\begin{table}

\caption{\label{tab:table}Comparison of the MSE of the ranking 
        estimators. SE stands for standard error.\label{tab:tab_sim_res}}
\centering
\begin{tabular}[t]{lrl}
\toprule
  & MSE & (Monte Carlo SE)\\
\midrule
Rank based on average & 4.310 & (0.071)\\
BR: normal outcome with homogeneous residuals & 3.715 & (0.055)\\
BR: ordinal outcome with homogeneous residuals & 3.807 & (0.056)\\
BR: normal outcome with heterogeneous residuals & 3.810 & (0.056)\\
BR: ordinal outcome with heterogeneous residuals & 3.517 & (0.054)\\
\bottomrule
\end{tabular}
\end{table}

\pagebreak

\hypertarget{discussion}{%
\section{\texorpdfstring{Discussion
\label{sec:disc}}{Discussion }}\label{discussion}}

Inspired by work on ranking baseball players
\citep{berger_bayesian-approach_1988}, health care facilities
\citep{lingsma_comparing_2010} and treatment effects
\citep{salanti_graphical_2011}, we developed a Bayesian Ranking method
based on a hierarchical model to support decision making on proposals
submitted to the SNSF. A provisional funding line is defined based on
the expected rank (ER) and the available budget. The ER and its credible
interval are then used to identify proposals with similar quality and
credible intervals that overlap with the provisional funding line. These
proposals are entered into a lottery to select those to be funded. The
approach acknowledges that there are proposals of similar quality and
merit, which cannot all be funded. Previous studies suggested that peer
review has difficulties in discriminating between applications that are
neither clearly competitive nor clearly non competitive
\citep{fang_research_2016, klaus_talent_2018, scheiner_predictive_2013}.
Decisions on these proposals typically lead to lengthy panel
discussions, with an increased risk of biased decision making. The
method proposed here avoids such discussions and thus may increase the
efficiency of the process, reduce bias and costs.

The Bayesian ranking methodology compares every assessor to every other
panel member. The ER considers all uncertainty in the evaluation process
that can be observed and quantified. It accommodates the fact that
different assessors have different grading habits. The model can also be
adjusted for potential confounding variables, such as external peer
reviewers' characteristics that influence their scores. A recent
analysis of 38'250 peer review reports on 12'294 SNSF project grant
applications across all disciplines showed that male reviewers, and
reviewers from outside Switzerland, awarded higher scores than female
reviewers and Swiss reviewers \citep{severin_gender_2020}.
\cite{Jayasinghe2003} modeled peer review records from the Australian
Research Council using a multilevel cross-classified model in order to
identify researcher and assessor attributes that influence the
evaluation. They found low reliability in the scores attributed by the
assessors, no gender effects, but an effect of the status of the
University and the academic rank of the first applicant.

We agree with \cite{goldstein_league_1996} who argued that ``no amount
of fancy statistical footwork will overcome basic inadequacies in either
the appropriateness or the integrity of the data collected''. In an
ideal world, all proposals would be evaluated by as many experts it
takes to ensure that meaningful differences between aggregated scores
can be detected with confidence. Evaluations would be unbiased and
describe nothing else but the quality of the proposals. Human nature and
limited resources regarding time and funding sadly prevent this ideal
situation from becoming a reality. The evaluation of grant proposals
will always be subjective to some extent and affected by unconscious
biases and chance. However, we are confident that the method presented
here is an improvement over the commonly used approaches to ranking
proposals and defining funding lines. Our approach should not be seen as
a mechanistic cookbook approach to decision making but as a tool that
can provide decision support for proposals of similar or
indistinguishable quality around the funding line. For example,
judgement continues to be required to decide whether a lottery should be
used or not.

We applied the approach to two instruments in career and project funding
at the SNSF. Our case studies addressed the specific context of the SNSF
and the two funding schemes and results may not be generalizable to
other instruments or funders. Further, we acknowledge that the team
carrying out this study included several researchers affiliated with the
SNSF. As the researchers' expectations might influence interpretation,
critical comment and review of our approach from independent scholars
and other funders will be particularly welcome.\\
In our first model, we treated the ordinal scores (from 1 to 6) as
continuous variables, thus assuming that the distance between each set
of subsequent scores is equal. This assumption might not always be
appropriate, but it builds on the currently used average-based ranking
method, as this procedure also assumes normality of the grades. We
showed how our initial model can be generalised to allow for an ordinal
outcome variable. However, because of the fact that for the ordinal
model extra parameters have to be estimated and in the simulation
studies this more general model did not always perform better, we
suggest to use the simpler model based on the normal likelihood.

The choice of priors in Bayesian models is always disputable. Especially
for the variance parameters, alternative prior distributions could be
investigated, like half-normal and half-Cauchy priors, whereas
inverse-Gamma priors are generally not recommended
\citep{gelman_prior_2006}. Instead of a Gibbs sampler, the
\texttt{brms}-package in \texttt{R} \citep{burkner_advanced_2018} can be
used, which makes the programming language Stan accessible with commonly
used mixed model syntax, similar to the \texttt{lme4}-package by
\cite{bates_fitting_2015}. Another approach is to use the \texttt{R}
code provided by \cite{lingsma_comparing_2010}. They implemented a
frequentist approach where the posterior means and variances of the
proposal-specific random intercept are approximated. However, if the
proposal effects are a posteriori dependent, because of the same
assessors evaluating the same set of proposals, the Bayesian approach is
easier to implement.

The use of a lottery to allocate research funding is controversial. At
the SNSF the applicants are informed about the possible use of random
selection, thus complying with the San Francisco Declaration on Research
Assessment \citep{noauthor_dora_2019}, which states that funders must be
explicit about assessment criteria. Of note, in the context of the
Explorer Grant scheme of the New Zealand Health Research Council,
\cite{liu_acceptability_2020} recently reported that most applicants
agreed with random selection. So far, the SNSF received no negative or
positive reactions to the use of random selection from applicants
\citep{bieri_how_2020}. The lottery element being controversial is also
the reason why we start using the Bayesian ranking methodology with a
50\% CrI. This ensures that the random selection groups are reasonably
small.

An alternative to ranks based on simple averages of scores is the
procedure employed by the National Institutes of Health (NIH). The NIH
uses percentile rankings to define a funding
line\footnote{see \url{https://www.niaid.nih.gov/grants-contracts/understand-paylines-percentiles}}:
proposals are first discussed and assigned scores from 1 to 9 by all
panel members. For each proposal, all scores are then averaged to obtain
an overall impact score. Finally, percentiles are determined by matching
overall impact scores against historical relative rankings (based on the
last three calls/review cycles). The NIH procedure makes a relative
ranking possible by putting the different scores in context, however,
when drawing the funding line (or payline), the uncertainty of the
estimated quantities is not taken into account. This is an important
difference to a ranking based on PCER (which is also percentile-based),
ER or SUCRA. Ignoring uncertainty is especially problematic since there
is evidence that the percentile scores employed by the NIH are poorly
predictive of grant productivity \citep{fang2016}.

In conclusion, we propose that a Bayesian modelling approach to ranking
proposals combined with a modified lottery can improve the evaluation of
grant and fellowship proposals. More research on the limitations
inherent in peer review and grant evaluation is needed. Funders should
be creative when investigating the merit of different evaluation
strategies \citep{severin_research_2020}. We encourage other funders to
conduct studies and test evaluation approaches to improve the evidence
base for rational and fair research funding.

\hypertarget{supplemental-materials-and-data}{%
\section*{Supplemental Materials and
Data}\label{supplemental-materials-and-data}}
\addcontentsline{toc}{section}{Supplemental Materials and Data}

An online fully reproducible supplement is provided which uses an
\texttt{R} (\texttt{ERforResearch}) package with the implementation of
the above presented methodology (see
\href{https://snsf-data.github.io/ERpaper-online-supplement/}{snsf-data.github.io/ERpaper-online-supplement/}).
The data used in the case studies can be downloaded from Zenodo:
\url{https://doi.org/10.5281/zenodo.4531160}.

\hypertarget{acknowledgment}{%
\section*{Acknowledgment}\label{acknowledgment}}
\addcontentsline{toc}{section}{Acknowledgment}

We are grateful to Hans van Houwelingen and Ewout Steyerberg for helpful
comments on an earlier version of this manuscript and to the National
Research Council of the SNSF for fruitful discussions. We also thank
Malgorzata Roos for further feedback on the manuscript as well as on the
implementation in \texttt{R}.

\newpage

\hypertarget{refs}{}
\begin{CSLReferences}{0}{0}
\end{CSLReferences}

\bibliography{ER_paper_lib}

\hypertarget{appendix}{%
\section*{Appendix}\label{appendix}}
\addcontentsline{toc}{section}{Appendix}

\setcounter{section}{0}
\renewcommand{\thesection}{\Alph{section}}

\hypertarget{analytical-formulas}{%
\section{\texorpdfstring{Analytical formulas
\label{app:analytical_stuff}}{Analytical formulas }}\label{analytical-formulas}}

This section will give some insights on how the previously discussed
quantities can be computed analytically rather than using MCMC samples.
The probability of \(\theta_i\) being smaller than \(\theta_k\), which
is used for the ER, can be computed as follows: \begin{eqnarray*}
\mbox{Pr}(\theta_i \leq \theta_k) & = & \mbox{Pr}(\theta_i - \theta_k \leq 0) \nonumber \\
& = & \mbox{Pr}\left (\frac{\theta_i - \theta_k - (\hat{\theta}_i - \hat{\theta}_k)}{\sqrt{\mbox{var}(\hat{\theta}_i) + \mbox{var}(\hat{\theta}_k) - 2 \mbox{cov}(\hat{\theta}_i, \hat{\theta}_k)}} \leq \frac{ - (\hat{\theta}_i - \hat{\theta}_k)}{\sqrt{\mbox{var}(\hat{\theta}_i) + \mbox{var}(\hat{\theta}_k) - 2 \mbox{cov}(\hat{\theta}_i, \hat{\theta}_k)}} \right) \nonumber \\
& = & \Phi \left (\frac{\hat{\theta}_k - \hat{\theta}_i}{\sqrt{\mbox{var}(\hat{\theta}_i) + \mbox{var}(\hat{\theta}_k) - 2\mbox{cov}(\hat{\theta}_i, \hat{\theta}_k)}} \right) , 
\label{eq:post-dist}
\end{eqnarray*} where \(\Phi\) is the standard normal cumulative
distribution function. Here, \(\hat{\theta}_i\) denotes the posterior
expectation of \(\theta_i\) and \(\mbox{var}(\hat{\theta}_i)\) and
\(\mbox{cov}(\hat{\theta}_i, \hat{\theta}_k)\) the corresponding
posterior variance and covariance. If the \(\theta_i\)'s are a
posteriori independent (which does in general not hold for model
\eqref{eq:final-model}), then
\(\mbox{cov}(\hat{\theta}_i, \hat{\theta}_k) = 0\) for \(i \neq k\).

There is also an analytical version of the posterior variance of the
rank \(R_i\) discussed in the Appendix of \cite{laird_empirical_1989}
that can be used to compute confidence intervals rather than credible
intervals. According to the latter authors, this posterior variance is
given by: \begin{eqnarray*}
\mbox{var}(R_i) & = & \mbox{var}\left(\sum_{k=1}^n \mbox{I} (\theta_i \leq \theta_k) \right) \nonumber \\
& = & \sum_{k=1}^n \mbox{var}\left(\mbox{I}(\theta_i \leq \theta_k) \right) + 2 \sum_{k=1}^n\sum_{l>k}\mbox{cov}\left(\mbox{I}(\theta_i \leq \theta_k), \mbox{I}(\theta_i \leq \theta_l)  \right) \nonumber \\
& = & \sum_{k=1}^n \mbox{Pr}(\theta_i \leq \theta_k ) \cdot (1 - \mbox{Pr}(\theta_i \leq \theta_k )) + \nonumber \\
& & \ \ \ \ \ 2 \cdot \sum_{k=1}^n \sum_{l>k} \big[ \mbox{Pr}(\theta_i \leq \mbox{min}(\theta_k, \theta_l) ) - \mbox{Pr}(\theta_i \leq \theta_k) \cdot \mbox{Pr}(\theta_i \leq \theta_l) \big]. 
\label{eq:var-er}
\end{eqnarray*}

\hypertarget{relationship-between-er-and-sucra}{%
\section{\texorpdfstring{Relationship between ER and SUCRA
\label{app:ersucra}}{Relationship between ER and SUCRA }}\label{relationship-between-er-and-sucra}}

In the following, the relationship between the SUCRA and the ER is
derived. Note that the ER, i.e.~the expectation of the rank, can be
expressed in terms of the rank probabilities as follows:
\(\mbox{ER}_i = \sum_{j=1}^n j \cdot \mbox{Pr}(i = j)\).
\begin{eqnarray*}
\mbox{SUCRA}_i & = & \frac{1}{n-1} \sum_{m=1}^{n-1} \mbox{cum}_{im}  = \frac{1}{n-1} \sum_{m=1}^{n-1} \sum_{j=1}^m \mbox{Pr}(i = j) \nonumber \\
(n - 1) \mbox{SUCRA}_i & = & \sum_{m=1}^{n-1} \sum_{j=1}^m \mbox{Pr}(i = j) \nonumber \\
(n - 1) \mbox{SUCRA}_i - n & = & \sum_{m=1}^{n-1} \sum_{j=1}^m \mbox{Pr}(i = j)  - \sum_{m=1}^n \sum_{j=1}^n \mbox{Pr}(i = j), \mbox{ because } \sum_{j=i}^n \mbox{Pr}(i = j) = 1\nonumber \\
- (n - 1) \mbox{SUCRA}_i + n & = & - \sum_{m=1}^{n-1} \sum_{j=1}^m \mbox{Pr}(i = j)  + \sum_{m=1}^n \sum_{j=1}^n \mbox{Pr}(i = j)\nonumber \\
n - (n - 1) \mbox{SUCRA}_i & = & \sum_{m=1}^{n-1} \sum_{j=m+1}^n \mbox{Pr}(i = j)  + \sum_{j=1}^n \mbox{Pr}(i = j)\nonumber \\
n - (n - 1) \mbox{SUCRA}_i & = & \sum_{j=1}^n \mbox{Pr}(i = j) + \sum_{j=2}^n \mbox{Pr}(i = j) + \dots + \sum_{j=n}^n \mbox{Pr}(i = j)\nonumber \\
n - (n - 1) \mbox{SUCRA}_i & = & \sum_{j=1}^n j \cdot \mbox{Pr}(i = j) = \mbox{ER}_i
\end{eqnarray*}

\hypertarget{implementation-of-the-bayesian-model-in-and-convergence-diagnostics}{%
\section{\texorpdfstring{Implementation of the Bayesian model in
\texttt{rjags} and convergence diagnostics
\label{app:conv}}{Implementation of the Bayesian model in  and convergence diagnostics }}\label{implementation-of-the-bayesian-model-in-and-convergence-diagnostics}}

The following code describes the definition of the model in \texttt{R}
through the package \texttt{rjags}. Note that this JAGS model definition
refers to the model described in \eqref{eq:merged-model}, exactly as it
is used in Section \ref{subsec:pf}. Find more alternative model
definitions in the function \texttt{get\_default\_jags\_model()} in our
package \texttt{ERforResearch}.

\begin{Shaded}
\begin{Highlighting}[]
\StringTok{"model\{}
\StringTok{      \# Likelihood:}
\StringTok{      for (i in 1:n) \{ \# i is not the application but the review}
\StringTok{          grade[i] \textasciitilde{} dnorm(mu[i], inv\_sigma2)}
\StringTok{                \# inv\_sigma2 is precision (1 / variance)}
\StringTok{          mu[i] \textless{}{-} overall\_mean + application\_intercept[num\_application[i]] +}
\StringTok{          voter\_intercept[num\_application[i], num\_voter[i]] +}
\StringTok{          section\_intercept[num\_section[i]]}
\StringTok{      \}}
\StringTok{      \# Ranks:}
\StringTok{      rank\_theta[1:n\_application] \textless{}{-} rank({-}application\_intercept[])}
\StringTok{      \# Priors:}
\StringTok{      for (j in 1:n\_application)\{}
\StringTok{        application\_intercept[j] \textasciitilde{} dnorm(0, inv\_tau\_application2)}
\StringTok{      \}}
\StringTok{      for (l in 1:n\_voters)\{}
\StringTok{        for(j in 1:n\_application)\{}
\StringTok{          voter\_intercept[j, l] \textasciitilde{} dnorm(nu[l], inv\_tau\_voter2)}
\StringTok{        \}}
\StringTok{      \}}
\StringTok{      for (l in 1:n\_voters)\{}
\StringTok{        nu[l] \textasciitilde{} dnorm(0, 4) }
\StringTok{      \}}
\StringTok{      for (l in 1:n\_section)\{}
\StringTok{        section\_intercept[l] \textasciitilde{} dnorm(0, inv\_tau\_section2)}
\StringTok{      \}}
\StringTok{      sigma \textasciitilde{} dunif(0.000001, 2)}
\StringTok{      inv\_sigma2 \textless{}{-} pow(sigma, {-}2)}
\StringTok{      inv\_tau\_application2 \textless{}{-} pow(tau\_application, {-}2)}
\StringTok{      tau\_application \textasciitilde{} dunif(0.000001, 2)}
\StringTok{      inv\_tau\_voter2 \textless{}{-} pow(tau\_voter, {-}2)}
\StringTok{      tau\_voter \textasciitilde{} dunif(0.000001, 2)}
\StringTok{      inv\_tau\_section2 \textless{}{-} pow(tau\_section, {-}2)}
\StringTok{      tau\_section \textasciitilde{} dunif(0.000001, 2)}
\StringTok{    \}"}
\end{Highlighting}
\end{Shaded}

Trace plots and the \(\widehat{R}\) values of the Gelman-Rubin
convergence diagnostic of the most important parameters can be computed;
see
\href{https://snsf-data.github.io/ERpaper-online-supplement/index.html}{Figure 3.1 and Table 3.1 of the online supplement}.
For reproducibility purposes, the length of burn-in and adaptations
phases, and the final number of iterations used in our case studies can
be retrievd from our online supplement as well as the seed used for the
sampling. With this information all models converged (with
\(\widehat{R}\) values lower to 1.1). Note that for convenience, we kept
the numbers of iterations high and did not try to optimize.
The online supplement also briefly discusses the sensitivity of the results to the choice of the hyperpriors; see \href{https://snsf-data.github.io/ERpaper-online-supplement/index.html}{Section 4 of the supplement}.

\hypertarget{code-simulating-data}{%
\section{\texorpdfstring{Code simulating data
\label{app:code}}{Code simulating data }}\label{code-simulating-data}}

Find the \texttt{R} code used for simulating the data in Section
\ref{sec:sim} below.

\begin{Shaded}
\begin{Highlighting}[]
\CommentTok{\# Simulate data afer Cao et al.:}
\FunctionTok{set.seed}\NormalTok{(}\DecValTok{0307}\NormalTok{)}
\NormalTok{J }\OtherTok{\textless{}{-}} \DecValTok{10}
\NormalTok{j }\OtherTok{\textless{}{-}} \DecValTok{1}\SpecialCharTok{:}\NormalTok{J}
\NormalTok{I }\OtherTok{\textless{}{-}} \DecValTok{50}
\NormalTok{N }\OtherTok{\textless{}{-}} \DecValTok{500}

\NormalTok{sims }\OtherTok{\textless{}{-}} \FunctionTok{list}\NormalTok{()}
\ControlFlowTok{for}\NormalTok{ (n }\ControlFlowTok{in} \DecValTok{1}\SpecialCharTok{:}\NormalTok{N)\{}
\NormalTok{  alpha }\OtherTok{\textless{}{-}} \FunctionTok{runif}\NormalTok{(J, }\SpecialCharTok{{-}}\NormalTok{.}\DecValTok{5}\NormalTok{, .}\DecValTok{5}\NormalTok{)}
\NormalTok{  sigma }\OtherTok{\textless{}{-}} \FunctionTok{runif}\NormalTok{(J }\SpecialCharTok{{-}} \DecValTok{1}\NormalTok{, }\DecValTok{0}\NormalTok{, .}\DecValTok{5}\NormalTok{)}
\NormalTok{  sigma }\OtherTok{\textless{}{-}} \FunctionTok{c}\NormalTok{(sigma, .}\DecValTok{9}\NormalTok{)}
  
\NormalTok{  c }\OtherTok{\textless{}{-}} \FunctionTok{rep}\NormalTok{(}\ConstantTok{NA}\NormalTok{, }\DecValTok{5}\NormalTok{)}
\NormalTok{  c[}\DecValTok{1}\NormalTok{] }\OtherTok{\textless{}{-}} \SpecialCharTok{{-}}\DecValTok{1}
\NormalTok{  c[}\DecValTok{2}\NormalTok{] }\OtherTok{\textless{}{-}} \SpecialCharTok{{-}}\NormalTok{.}\DecValTok{5}
\NormalTok{  c[}\DecValTok{3}\NormalTok{] }\OtherTok{\textless{}{-}} \DecValTok{0}
\NormalTok{  c[}\DecValTok{4}\NormalTok{] }\OtherTok{\textless{}{-}}\NormalTok{ .}\DecValTok{5}
\NormalTok{  c[}\DecValTok{5}\NormalTok{] }\OtherTok{\textless{}{-}} \DecValTok{1}
  
\NormalTok{  theta }\OtherTok{\textless{}{-}} \FunctionTok{rnorm}\NormalTok{(I, }\DecValTok{0}\NormalTok{, }\DecValTok{1}\NormalTok{)}
  
\NormalTok{  x }\OtherTok{\textless{}{-}}\NormalTok{ y }\OtherTok{\textless{}{-}} \FunctionTok{matrix}\NormalTok{(}\ConstantTok{NA}\NormalTok{, }\AttributeTok{nrow =}\NormalTok{ I, }\AttributeTok{ncol =}\NormalTok{ J)}
  \ControlFlowTok{for}\NormalTok{ (ii }\ControlFlowTok{in} \DecValTok{1}\SpecialCharTok{:}\NormalTok{I)\{}
    \ControlFlowTok{for}\NormalTok{ (jj }\ControlFlowTok{in} \DecValTok{1}\SpecialCharTok{:}\NormalTok{J)\{}
\NormalTok{      x[ii, jj] }\OtherTok{\textless{}{-}} \FunctionTok{rnorm}\NormalTok{(}\DecValTok{1}\NormalTok{, alpha[jj] }\SpecialCharTok{+}\NormalTok{ theta[ii], sigma[jj]}\SpecialCharTok{**}\DecValTok{2}\NormalTok{)}
\NormalTok{      y[ii, jj] }\OtherTok{\textless{}{-}} \DecValTok{1} 
      \ControlFlowTok{if}\NormalTok{ ((x[ii, jj] }\SpecialCharTok{\textgreater{}}\NormalTok{ c[}\DecValTok{1}\NormalTok{]) }\SpecialCharTok{\&}\NormalTok{ (x[ii, jj] }\SpecialCharTok{\textless{}=}\NormalTok{ c[}\DecValTok{2}\NormalTok{])) y[ii, jj] }\OtherTok{\textless{}{-}} \DecValTok{2}
      \ControlFlowTok{if}\NormalTok{ ((x[ii, jj] }\SpecialCharTok{\textgreater{}}\NormalTok{ c[}\DecValTok{2}\NormalTok{]) }\SpecialCharTok{\&}\NormalTok{ (x[ii, jj] }\SpecialCharTok{\textless{}=}\NormalTok{ c[}\DecValTok{3}\NormalTok{])) y[ii, jj] }\OtherTok{\textless{}{-}} \DecValTok{3}
      \ControlFlowTok{if}\NormalTok{ ((x[ii, jj] }\SpecialCharTok{\textgreater{}}\NormalTok{ c[}\DecValTok{3}\NormalTok{]) }\SpecialCharTok{\&}\NormalTok{ (x[ii, jj] }\SpecialCharTok{\textless{}=}\NormalTok{ c[}\DecValTok{4}\NormalTok{])) y[ii, jj] }\OtherTok{\textless{}{-}} \DecValTok{4}
      \ControlFlowTok{if}\NormalTok{ ((x[ii, jj] }\SpecialCharTok{\textgreater{}}\NormalTok{ c[}\DecValTok{4}\NormalTok{]) }\SpecialCharTok{\&}\NormalTok{ (x[ii, jj] }\SpecialCharTok{\textless{}=}\NormalTok{ c[}\DecValTok{5}\NormalTok{])) y[ii, jj] }\OtherTok{\textless{}{-}} \DecValTok{5}
      \ControlFlowTok{if}\NormalTok{ (x[ii, jj] }\SpecialCharTok{\textgreater{}}\NormalTok{ c[}\DecValTok{5}\NormalTok{]) y[ii, jj] }\OtherTok{\textless{}{-}} \DecValTok{6}
\NormalTok{    \}}
\NormalTok{  \}}
  
  \CommentTok{\# How many NAs/COIs do we want to introduce?}
\NormalTok{  n\_nas }\OtherTok{\textless{}{-}} \FunctionTok{sample}\NormalTok{(}\DecValTok{0}\SpecialCharTok{:}\DecValTok{100}\NormalTok{, }\DecValTok{1}\NormalTok{)}
  \CommentTok{\# Which elements of the matrix will be NA? }
  \CommentTok{\# Make sure, we did not set a whole column or row to NA}
\NormalTok{  test }\OtherTok{\textless{}{-}} \ConstantTok{TRUE}
  \ControlFlowTok{while}\NormalTok{ (test)\{}
\NormalTok{    where\_nas }\OtherTok{\textless{}{-}} \FunctionTok{sample}\NormalTok{(}\DecValTok{1}\SpecialCharTok{:}\NormalTok{(I}\SpecialCharTok{*}\NormalTok{J), n\_nas)}
\NormalTok{    xx }\OtherTok{\textless{}{-}}\NormalTok{ x; yy }\OtherTok{\textless{}{-}}\NormalTok{ y}
\NormalTok{    xx[where\_nas] }\OtherTok{\textless{}{-}} \ConstantTok{NA}
\NormalTok{    yy[where\_nas] }\OtherTok{\textless{}{-}} \ConstantTok{NA}
\NormalTok{    test }\OtherTok{\textless{}{-}} \FunctionTok{any}\NormalTok{(}\FunctionTok{rowSums}\NormalTok{(}\FunctionTok{is.na}\NormalTok{(xx)) }\SpecialCharTok{==}\NormalTok{ J) }\SpecialCharTok{|} \FunctionTok{any}\NormalTok{(}\FunctionTok{colSums}\NormalTok{(}\FunctionTok{is.na}\NormalTok{(xx)) }\SpecialCharTok{==}\NormalTok{ I)}
\NormalTok{  \}}
  \CommentTok{\# Creation of data in list format}
\NormalTok{  sims[[n]] }\OtherTok{\textless{}{-}} \FunctionTok{list}\NormalTok{(}\AttributeTok{x =}\NormalTok{ xx, }\AttributeTok{y =}\NormalTok{ yy, }\AttributeTok{theta =}\NormalTok{ theta, }\AttributeTok{rank =} \FunctionTok{rank}\NormalTok{(}\SpecialCharTok{{-}}\NormalTok{theta))}
\NormalTok{\}}
\end{Highlighting}
\end{Shaded}

\end{document}